# Mode-selective ballistic pathway to a metastable electronic phase

Hannes Böckmann[1,2*], Jan Gerrit Horstmann[1,2], Abdus Samad Razzaq[3], Stefan Wippermann[3*], and Claus Ropers[1,2]

[1]Max Planck Institute for Multidisciplinary Sciences, Göttingen 37077, Germany.
[2]4th Physical Institute, Solids and Nanostructures, University of Göttingen, Göttingen 37077, Germany.
[3]Max-Planck-Institut für Eisenforschung GmbH, Düsseldorf 40237, Germany.

**Exploiting vibrational excitation for the dynamic control of material properties is an attractive goal with wide-ranging technological potential [1–4]. Most metal-to-insulator transitions are mediated by few structural modes and are thus ideal candidates for the selective driving towards a desired electronic phase. Such targeted navigation within a generally multi-dimensional potential energy landscape requires microscopic insight into the non-equilibrium pathway. However, the exact role of coherent inertial motion across the transition state has remained elusive. Here, we demonstrate mode-selective control over the metal-to-insulator phase transition of atomic indium wires on the Si(111) surface, monitored by ultrafast low-energy electron diffraction. We use tailored pulse sequences to individually enhance or suppress key phonon modes and thereby steer the collective atomic motion within the potential energy surface underlying the structural transformation. *Ab initio* molecular dynamics simulations demonstrate the ballistic character of the structural transition along the deformation vectors of the Peierls amplitude modes. Our work illustrates that coherent excitation of collective modes via exciton-phonon interactions evades entropic barriers and enables the dynamic control of materials functionality.**

## Introduction

The interactions of individual particles govern the transition from the atomistic structure of matter to the vast variability of materials properties, with a microscopic description challenged by the complexity of many-body physics. Yet, for many particularly interesting phenomena, such as emergent states and phases, the coordinated motion of many microscopic degrees of freedom allows for a drastic reduction of dimensionality into few collective modes [5]. Correlated systems often feature a strong coupling between electronic and structural degrees of freedom, in which the balance between nearby phases is highly susceptible to an optical stimulus [3,6–12]. Studying coherent structural dynamics in such systems thus offers possibilities for uncovering unifying concepts of cooperative action and using light as a tool to manipulate matter at the nanoscale.

The metal-to-insulator transition [13] in charge density wave materials is an intriguing example, as the accompanying periodic lattice distortion is governed by specific phonon modes [14,15]. In particular, the Peierls system of atomic indium chains on a Si(111) surface [16] has attracted substantial interest due to the ultrafast nature of the transformation in response to femtosecond optical excitation [17–23]. At high excitation densities, the rapid phase transition between the insulating (8×2) and the metallic (4×1) states is explained by the driven motion within a strongly deformed potential energy surface (PES) [17]. Close to the excitation threshold of the phase change, however, recent experimental work implies relevant contributions from vibrational coherence in both shear and rotational modes [18], which affect the outcome of the laser-triggered phase transformation [24,25]. These observations raise the question whether over-the-barrier transitions carried by coherent nuclear motion are a viable strategy for structural control.

Here, we present a combined experimental and theoretical study which demonstrates the steering of this system in a cooperative inertial manner through the underlying PES, exploiting vibrational coherences, and resulting in a metastable electronic phase. Experimentally, we employ an optical multi-pulse control scheme to repetitively excite the shear and rotation modes. We exert control over the vibrational amplitude via transient excitation of the PES and monitor the resulting impact on the phase transition efficiency using ultrafast low-energy electron diffraction (ULEED). By directing the trajectory along the individual modes towards the (4×1) structure, we retrieve a characteristic vibrational response which we can link to the ultrafast motion within the underlying PES. We perform density functional theory calculations to reveal the two-dimensional PES and the location of the transition state. Using *ab initio* molecular dynamics simulations, we demonstrate the decisive role of kinetic energy in traversing an off-diagonal transition state and overcoming the potential barrier.

## Conceptual framework for mode-selective driving of vibrational coherence

We first illustrate some characteristics of the studied system and the scheme for controlling its transition. At room temperature, indium atoms arrange in a zigzag pattern on the (111) face of silicon and induce a (4×1) superstructure [26]. Below the critical temperature of 125 K, a first-order Peierls transition [27] occurs between the metallic (4×1) and the insulating (8×2) phase, which manifests in the formation of a hexagonal bonding motif (Fig. 1a, left) [16,28,29]. A metastable (4×1) structure is, however, prepared by the sudden quenching of the charge density wave in response to irradiation with femtosecond optical pulses (Fig. 1a, right) [17,19–21].

For sequential subthreshold excitations, vibrational coherences in two key modes have significant influence on the transition efficiency (Fig. 1a, center) [18]. This observation strongly suggests that the light-induced structural transformation can be described in a configuration space spanned by these select collective modes (Fig. 1b (1)). Multi-pulse sequences adapted to the vibrational periods promise mode-selective control over the coherent structural evolution (Fig. 1b (2)) [30,31] and the optically induced switching probability as a function of the momentary displacements (Fig. 1b (3)).

## Experiment: Selective Control Probed by Ultrafast LEED

In the experiments, we implement such a multi-pulse excitation scheme and probe the resulting structural change by ultrafast LEED (Fig. 2a). The use of time-resolved low-energy electron diffraction allows for direct probing of the surface atomic structure and enables fundamental insight into non-equilibrium lattice dynamics [32–34] as well as structural phase transitions [18]. Two preceding weak "preparation pulses" (mutual delay $\tau$) induce a well-defined vibrational state, followed by a stronger "switch pulse" at delay t completing the transition. The transformed surface fraction is subsequently probed by an electron pulse, at a delay time $t_{p-el}$=75 ps after the switch pulse. The phase transition is accompanied by a suppression of (8×2) spots in the LEED pattern, thereby indicating the switched fraction of In chains as a function of the pulse-sequence timing (Fig. 2b).

Both optical delays are scanned in a cartesian grid to obtain a two-dimensional (2D) representation of the relative switching yield $Y(\tau, t)$ (Fig. 3a). Two main contributions dominate the 2D map. The coherent vibrational motion induced by the first and second preparation pulse causes superimposed diagonal and horizontal streaks [35]. Diagonal components arise from the combined response of both pulses while horizontal components follow from motion induced by the second pulse.

Figure 3b displays two selected traces as a function of the switch pulse timing, for the system prepared resonantly at the rotation (top: τ=1.2 ps) and shear (bottom: τ=1.9 ps) mode periods. Notably, the traces show prominent oscillations dominated by the respective mode. The resulting enhancement of the vibrational amplitudes is quantified via Fourier transformation of the individual traces (Fig. 3b insets). We find strongly asymmetric line shapes for both the shear mode and (to a smaller extent) the rotation mode, which we attribute to time-dependent electronic softening and anharmonicity of the excited PES. Gradual softening of amplitude modes is a common feature in Peierls systems close to threshold excitation [36,37].

In order to further explore the selectivity of the demonstrated multi-pulse control, we display the individual mode amplitudes as a function of preparation period τ. Phonon amplitudes are extracted from a one-dimensional Fourier transform along the response period t, $|F_{1D}(Y(\tau, t))(\tau, f_t)|$ (Fig. 3c). Overall, the switching efficiency is dominated by the shear mode. The rotation mode only prevails close to in-phase excitation and simultaneous out-of-phase excitation with the shear mode. An exclusive selection of shear mode displacement at complete suppression of rotation motion is achieved at τ=0.6 ps. The opposite case is not realized in this way, where optimized rotation mode excitation (τ~1.2 ps) leaves a remaining shear mode amplitude.

A complete picture of the coherent system response and the correlation between vibrational modes is obtained from a two-dimensional Fourier transform $|FT_{2D}(Y(\tau, t))(f_\tau, f_t)|$ (Fig. 3d), in close analogy to measurement schemes used in ultrafast vibrational spectroscopy [38]. The 2D spectrum is dominated by diagonal peaks at the shear and rotation frequencies, again demonstrating the mode-selective coherent control via repetitive displacive excitation of these phonons [39–41]. The shear mode peak is noticeably shifted from the diagonal ($f_\tau$=0.63 THz; $f_t$ =0.57 THz). This shift is ascribed to the increased electronic softening from accumulated excitation in the multi-pulse sequence, which causes the switch pulse to probe the system at a substantially lower frequency. Moreover, a higher-frequency (symmetric) shear mode,which is also observed in Raman and transient reflectivity measurements [18,42], may further contribute to the blue-shifted excitation. While the rotation mode peak is nearly centered on the diagonal, it exhibits a broadening along the horizontal direction, with substantial contributions also spanning to higher preparation frequencies.

Besides the diagonal peaks, we identify a cross peak at rotation preparation and shear response frequencies (bottom-right quadrant in Fig. 3d). The appearance of a cross peak is indicative for anharmonic coupling between the two modes, hence corroborating a description of the transformation within a 2D PES. An interesting observation is the missing opposite feature, which indicates a unidirectional coupling from the rotation to the shear mode. In addition, the asymmetric lineshape of the shear mode (Fig. 3b, inset) suggests that a Fano-type interference of the resonant contribution with the broad non-resonant electronic background may enhance the higher-energy lobe.

## Theory: Two-Dimensional Potential Energy Surface and Ultrafast Dynamics

In order to further elucidate the exact role of vibrational coherence for the optically-induced transformation, to shed light on the underlying PES, and to locate the transition state, we performed constrained density-functional theory (DFT) calculations and *ab initio* molecular dynamics (AIMD) simulations, using the amplitude mode displacements as reaction coordinates.

The (4×1) structure consists of two coupled Peierls chains [43]. Via the rotation mode, a period doubling of the translational symmetry is introduced that creates new *intrachain* covalent bonds (trimers) within each of the two chains and thereby opens a band gap close to the Brillouin zone boundary at $k_x = 0.5$ (Fig. 4a) [24,44]. The shear mode laterally displaces the two coupled chains with respect to each other [29], giving rise to new covalent *interchain* bonds [45] in the hexagonal bonding motif of the (8×2) structure. Here, a gap is opened at the zone center at $k_x = 0.0$ (Fig. 4a), since the translational symmetry remains unchanged. Optical excitation at $k_x = 0.0$ and $k_x = 0.5$ depopulates (populates) the bonding (antibonding) electronic states, associated to the covalent hexagon and trimer bonds formed upon shear and rotation displacements, respectively. In consequence, the electronic rearrangement weakens these bonds and thereby exerts forces between and within the two coupled indium chains that initiate the atomic shear and rotation motion, respectively (see Extended Figures S1) [16,24].

The resulting PES is obtained from partitioning the (8×2) → (4×1) transformation vector into shear and rotation contributions (Fig. 4b) (see Methods: Density Functional Theory). Apart from the stable (8×2) and metastable (4×1) structures, we identify a third (although unstable) configuration via a barrierless unshearing of the (8×2) structure, which is equivalent to the previously suggested trimer model [46,47] for the (8×2) low temperature ground state. These results clearly demonstrate that the minimum energy path traverses an off-diagonal transition state, as recently hypothesized [18]. The microscopic origin of the transition state location is its neighborhood to the energetically favorable trimer state. Hence, avoiding large barriers in the structural transformation first requires an unshearing, followed by a de-rotation.

We next discuss the PES in the presence of electronic excitations. For this purpose, we follow the numerical treatment suggested in previous studies [17,21], incorporating hot-electron populations of photoexcited charge carriers in the lowest conduction states at the zone center and highest valence states at the zone boundary (see Methods: Density Functional Theory). Upon population (depopulation) of the zone-center conduction (zone-boundary valence) states, the (8×2) structure becomes less favorable, and the potential minimum is displaced towards the (4×1) state (Fig. 4c). This excitation leaves the structure in a non-equilibrium configuration, the response to which we track by AIMD on the excited state PES. We note that the trajectories displayed in Fig. 4c are obtained by projecting the displacement of the nuclei on the shear and rotation transformation vectors. Only the electronic occupations are fixed, whereas the nuclear atomic motion is unconstrained.

Figure 4c describes the coherent atomic motion, arising from displacive excitation. The shown trajectories result from pulse sequences corresponding to a shear-mode preparation, and a response matched to either the shear or rotation period. It is evident that the acquired kinetic energy determines the final structural state after the excitation sequence. An in-phase excitation with the shear mode induces the transition (red trajectory), while an out-of-phase excitation may not (blue trajectory). We thus conclude that the kinetic energy is a decisive parameter in tilting the subtle balance towards one resulting phase, as the system is driven near a complete vanishing of the potential barrier. The results provide strong evidence for the directed collective atomic motion during the phase transition and confirm the existence of a ballistic control mechanism. Moreover, our AIMD simulations demonstrate that the induced nuclear motion is dominated by these two modes, before the acquired kinetic energy is dissipated into the reservoir of further phononic degrees of freedom. A persistent occupation for several picoseconds is also suggested by previous studies, where the discrepancy of timescales between electronic cooling in relevant band structure segments (less than 1 ps) [22] and lattice heating (~6 ps) [17] is rationalized by the preferential population of strongly coupled phonon modes.

We note that for an individual multi-pulse sequence, as simulated in our AIMD, the outcome of the switching process is discrete. Experimentally, in contrast, the switching probability is continuous and measured as an ensemble average, incorporating thermally induced fluctuations in the nuclear motion and varying barrier heights [18]. As comprehensive statistical sampling in AIMD is prohibitively expensive, we condense the full AIMD to a coarse-grained MD model and treat coupling to other phonon modes implicitly (see Methods: Coarse-Grained Molecular Dynamics). We find that the off-diagonal transition state and the corresponding asymmetry of the PES translate into characteristic timing-dependent amplitudes (Fig. 4d), in excellent qualitative agreement with the experiments (Fig. 3c). The modulation of the shear amplitude with $\tau$, however, is slightly stronger in the simulation compared to the experiment, with a ratio of 3:1 between maximum and minimum in the simulation compared to 2:1 in the experiment. This results in a somewhat higher (and narrower) peak of the shear mode in the simulated 2D spectrum (Fig. 4e), likely caused by inhomogeneous broadening from a distribution of barrier heights, which is not contained in the model.

In agreement with the experimental spectrum, the shear-mode peak in the simulated 2D spectrum also exhibits an off-diagonal side lobe. However, its location along $f_\tau$ does not coincide with the rotation frequency, suggesting that in the coarse-grained model, it does not arise from anharmonic couplings but rather from the above-mentioned interference of the resonance with the spectrally broad electronic response. This is an indication that the hot carrier relaxation dynamics play an important role for the photoinduced structural dynamics [22]. Including explicit *ab initio* derived relaxation dynamics of the electron and hole populations in the coarse-grained model, e.g. based on lifetimes obtained from many-body perturbation theory in GW approximation, may also yield insights into the microscopic origin of the apparent directional coupling from the rotation to the shear mode observed in the experiments.

## Conclusion

In a combined theoretical and experimental approach, we disentangle the roles of specific collective modes during the ultrafast (8×2) to (4×1) transformation in indium atomic wires and shed light on the underlying two-dimensional PES. Our results demonstrate the ballistic control over a structural phase transition via inducing mode-selective vibrational coherences by means of tailored optical pulse sequences, in close analogy to the guiding of reaction pathways in femtochemistry.

In future studies, the specificity in selecting vibrational excitations and shaping the PES could be further enhanced by tailoring the excitation to particular optical transitions in the band structure, exploring possible hidden phases or deterministically generating topological states. The presented scheme can further be generalized to diverse phenomena in which entropic barriers are circumvented by the directed motion along collective modes. Selective excitation of such modes may directly relate specific degrees of freedom to emerging properties and functionality. Beyond exotic phases in strongly correlated materials, future applications span from photo-induced molecular isomerization over protein functionality to quantum cellular automata.

## Figures

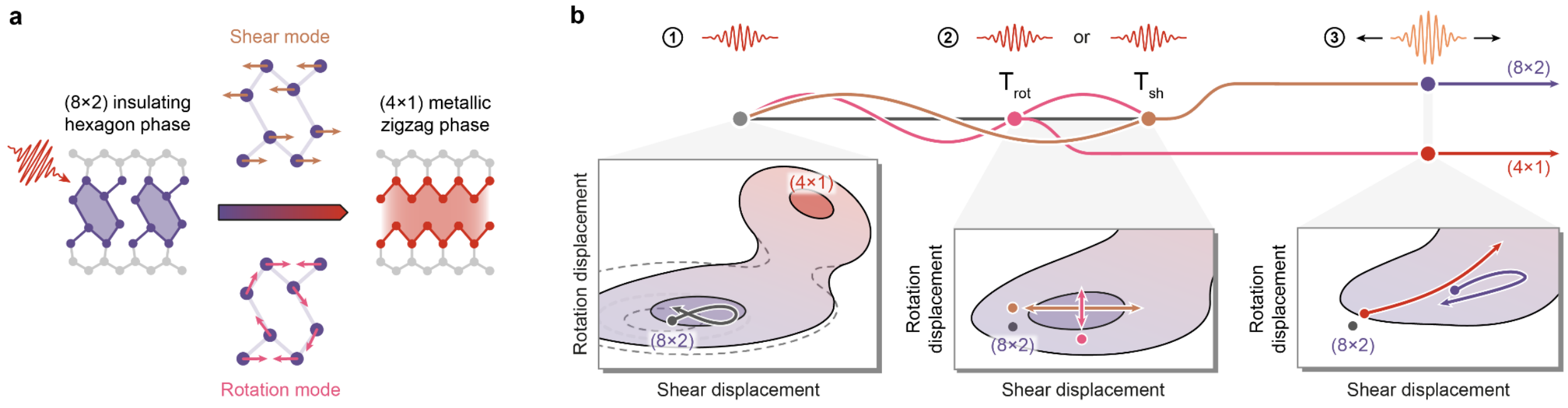

**Fig. 1:** **a**, Atomic indium wires undergo a structural phase transition upon optical excitation, mediated by distinct vibrational modes. **b**, Mode-selective steering of the ballistic trajectory. A first pulse initiates nuclear motion by displacive excitation of coherent phonons in the underlying 2D PES (1). The described trajectory is directed along the individual coordinates by timing a second pulse to the oscillation period (2). A third stronger switch pulse probes the impact of the momentary mode displacements on the switching efficiency (3).

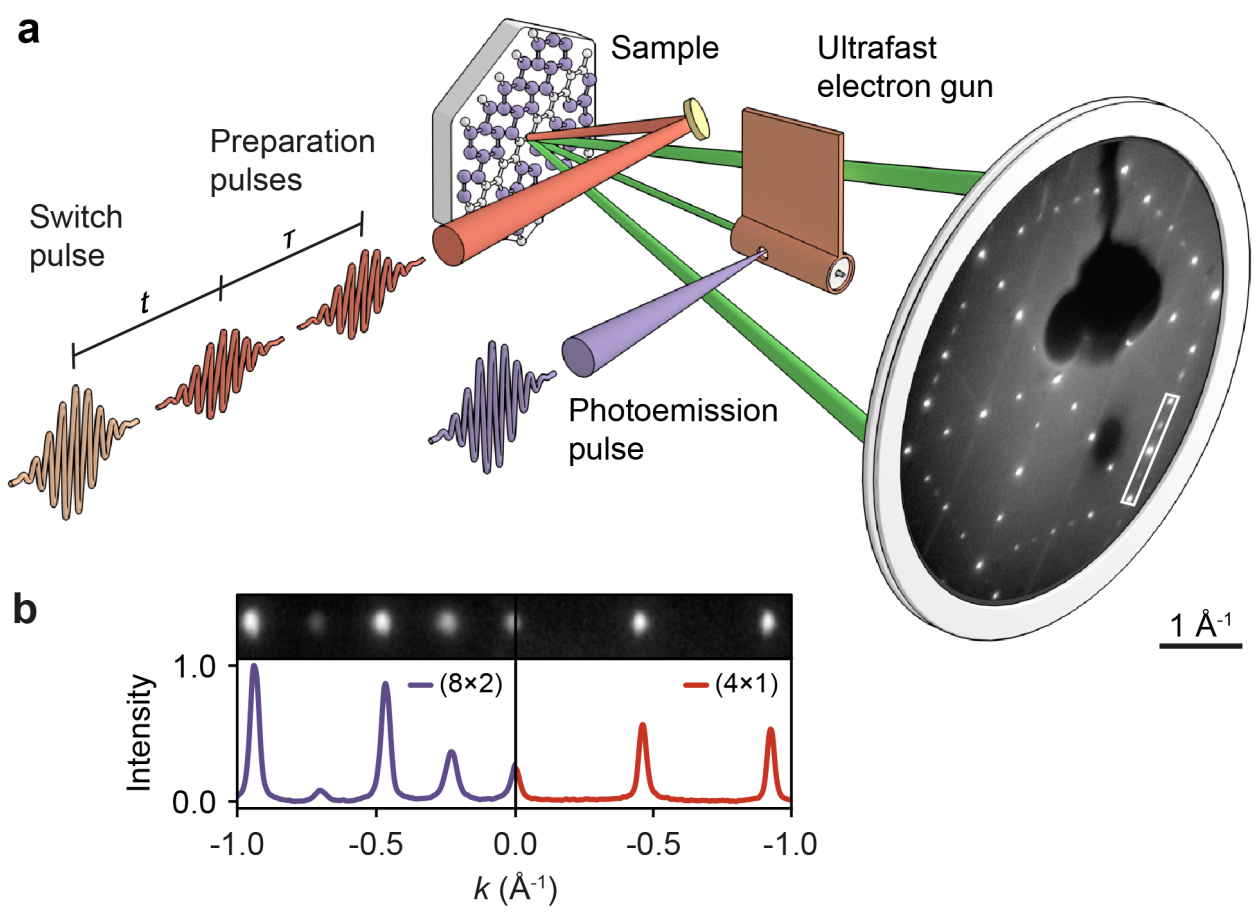

**Fig. 2:** **a**, Experimental scheme. Ultrashort electron pulses from a miniaturized laser-driven electron gun probe the microscopic structure of atomic indium wires on the Si(111) surface after optical excitation by a pulse train (preparation pulses: $\lambda$=800 nm, F=0.09 mJ/cm$^2$; switch pulse: $\lambda$=1030 nm, F=0.66 mJ/cm$^2$) in a LEED experiment. **b**, LEED line profiles for the insulating (8×2) and metallic (4×1) phase.

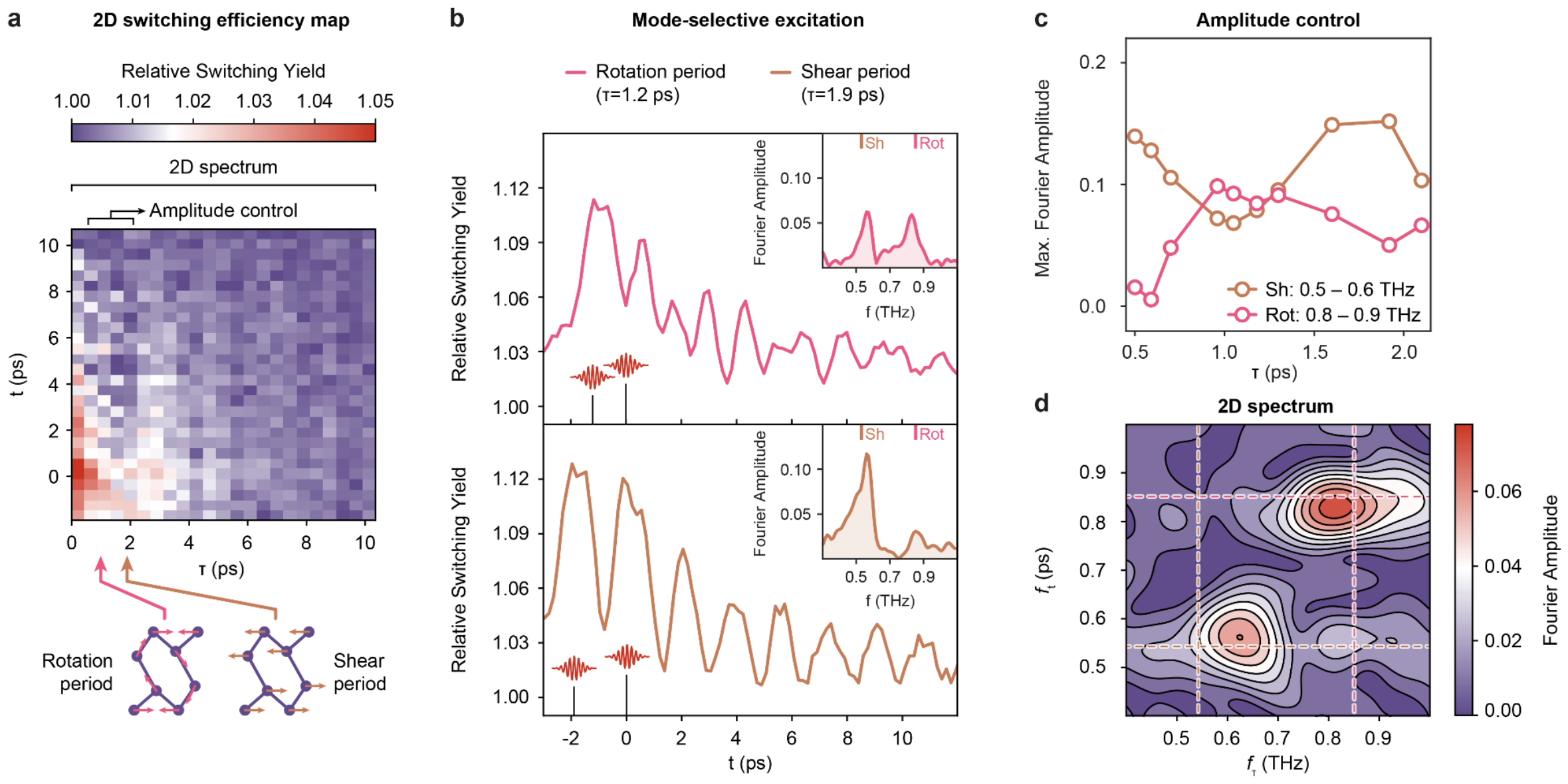

**Fig. 3:** **a**, Relative switching yield probed in a 2D scheme. Preparation period τ and response period t are scanned in a cartesian grid. **b**, High-resolution cuts along t, while τ is matched to either rotation (top) or shear (bottom) mode. Insets show Fourier transforms of the traces. Shear (0.54 THz) and rotation (0.85 THz) mode amplitudes are modulated via phase-matching. Low-frequency tails indicate electronic softening within the excited PES. **c**, τ-dependent modulation of shear (0.5 - 0.6 THz) and rotation (0.8 - 0.9 THz) mode amplitudes after Fourier transformation along t. **d**, 2D Fourier transform along τ and t. Prominent peaks appear at the rotation (pink, dashed) and shear (gold, dashed) mode frequencies.

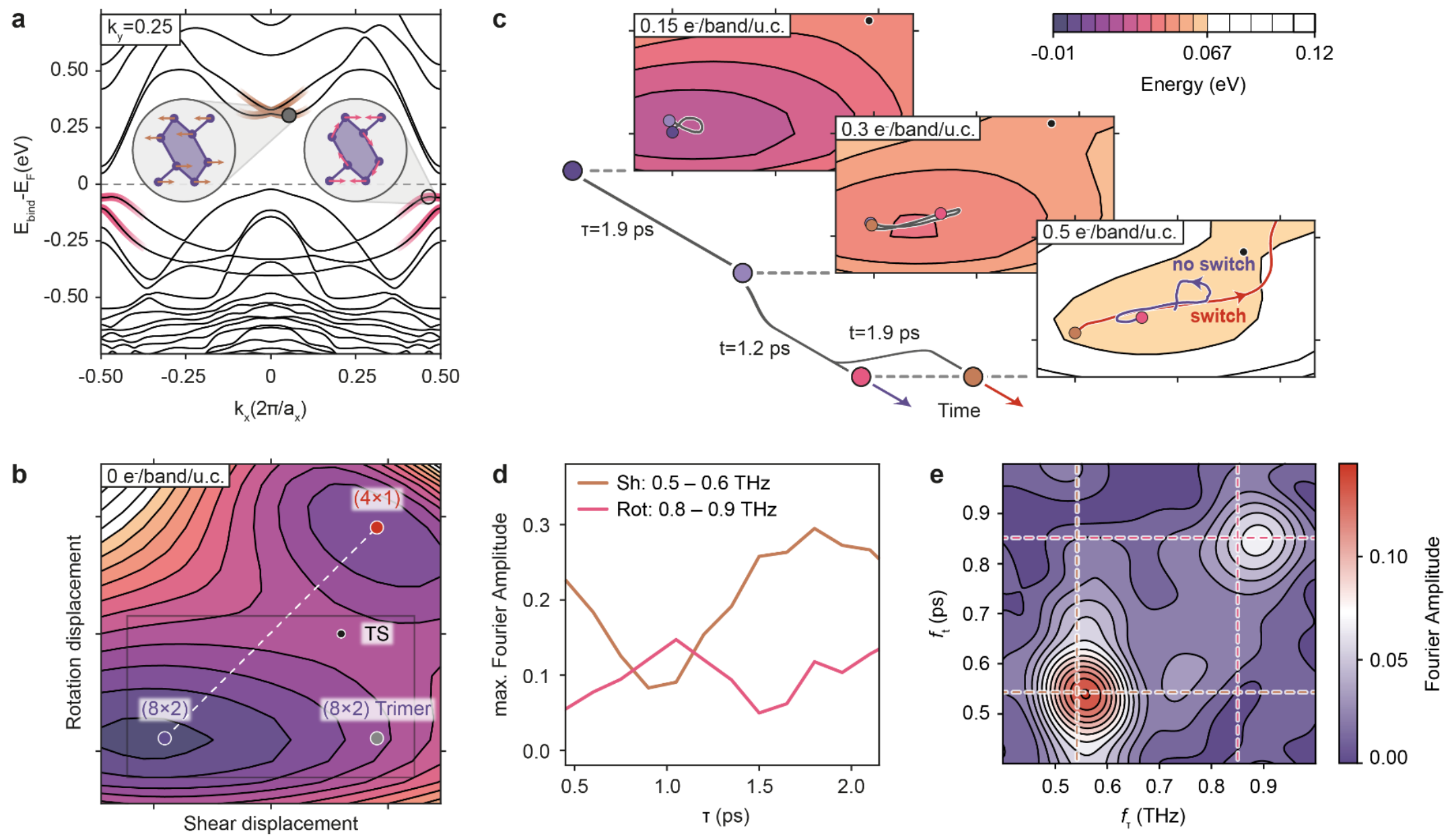

**Fig. 4:** **a**, Calculated electronic band structure of the Si(111)(8×2)-In phase. $k_x$ and $a_x$ denote the reciprocal and real space lattice vectors in wire direction. Electronic occupation of coloured segments couples strongly to atomic shear (gold) and rotation (pink) motion [25]. **b**, Calculated two-dimensional PES along the (8×2) → (4×1) phase transition as function of shear and rotation displacements shows an off-diagonal transition state (TS), shifted towards the (8×2) trimer state. **c**, PES segment with an increasing electronic occupation of surface bands (indicated in **a**). *Ab initio* molecular dynamics calculations yield the coherent motion, initiated by displacive excitation, for specific excitation sequences. The pulse timing directly determines the final state of the structure. **d**, τ-dependent mode amplitudes, extracted from coarse-grained MD simulations. **e**, Simulated 2D spectrum.

## Citations

## Acknowledgements

This work was funded by the European Research Council (ERC Starting Grant ‘ ULEED’, ID: 639119) and the Deutsche Forschungsgemeinschaft (SFB-1073, project A05). We gratefully acknowledge insightful discussions with Felix Kurtz.

## Author Contributions

The project was conceived by C.R., with contributions from H.B. Measurements and data analysis were performed by H.B., with contributions from J.G.H. Density functional theory and AIMD simulations were conceived by S.W., and conducted by A.S.R. and S.W. Coarse-grained MD simulations were conceived by S.W. and H.B., and carried out by H.B. The manuscript was written by H.B., J.G.H., S.W. and C.R. All authors discussed the results and commented on the manuscript.

## Methods

### Ultrafast LEED and optical set-up

ULEED is a novel tool for the investigation of structural dynamics at solid-state surfaces by means of an optical-pump/electron-probe scheme [18,32–34]. The technique combines the high surface sensitivity of low-energy electrons backscattered from a sample with laser-triggered electron sources and ultrafast optical excitation to follow the evolution of surface structure far from equilibrium.

To enable ULEED measurements with high temporal and momentum resolution, we have developed a miniaturized, laser-driven electron gun consisting of a nanometric tungsten tip as well as four metal electrodes (outer diameter 2 mm; for further details, see Refs. [18,33]). Ultrashort electron pulses are generated via localized two-photon photoemission by illuminating the tip apex with femtosecond laser pulses (central wavelength $\lambda_c$ = 400 nm, pulse duration $\Delta\tau$ = 45 fs, pulse energy $E_p$ = 20 nJ) from a noncollinear optical parametric amplifier at a repetition rate of 100 kHz. The small diameter of the electron gun assembly allows for distances of only a few millimeters between the electron source and the sample at a reasonably small fraction of shadowed electron diffraction signal. By decreasing the propagation length, the dispersion-induced broadening effect on the electron pulse is reduced, resulting in pulse durations down to 16 ps. The backscattered electrons from the surface are amplified and recorded by a combination of a chevron microchannel plate, a phosphor screen and a cooled sCMOS (scientific complementary metal–oxide–semiconductor) camera. The integration time for a single data point ranges from $t_{int} = 30 \times 1.5$ s = 45 s in Fig. 3b to $t_{int} = 30 \times 0.8$ s = 24 s in Fig. 3d.

In the ULEED experiments, the surface is excited by a pair of weak preparation pulses ($\lambda_c$ = 800 nm, $\hbar\omega$ = 1.55 eV, $\Delta\tau$ = 232 fs, focus area: (249±7)µm × (400±10)µm (FWHM)) generated from the output of an optical parametric amplifier in a Michelson interferometer, and a stronger switch pulse ($\lambda_c$ = 1,030 nm, $\hbar\omega 1$ = 1.2 eV, $\Delta\tau$ = 212 fs, focus area: (223±12)µm × (296±14)µm (FWHM)) from an Yb:YAG amplifier system. The mutual delay between the 800 nm pulse pair and the switch pulse at 1030 nm was controlled by a motorized linear delay stage. The final state of the surface is probed by 80 eV electron pulses ($\Delta\tau$ = 50 ps) at a fixed delay $\Delta t_{p-el}$=75 ps. The electron beam diameter (≈ 80 × 80 μm² FWHM) averages over a large ensemble of local configurations. Additional details about the ultrafast LEED and the optical setups are outlined in Ref. [18].

## Sample preparation

All experiments in this work have been conducted under ultra-high-vacuum conditions (base pressure $p < 2 \times 10^{-10}$ mbar). This minimizes the effects of adsorbates on the (8 × 2) ground state structure and, consequently, on the (8 × 2) → (4 × 1) transition [19,27]. The silicon substrates (phosphorus-doped wafers with resistivity R = 0.6–2 Ω cm) were cleaned by flash-annealing at $T_{max}$ = 1250 °C for five seconds through direct current heating (maximum pressure during the flashing procedure was kept below $p_{max} = 2 \times 10^{-9}$ mbar). In a next step, 1.2 monolayers of indium were deposited onto the resulting Si(111)(7 × 7) surface reconstruction at room temperature (deposition rate: $\Gamma_d$=0.5 ML/min) and annealed at T = 500 °C for 300 s. The resulting Si(111)(4 × 1)-In phase was inspected in our ultrafast LEED set-up and subsequently cooled to a base temperature of T = 60 K using an integrated continuous-flow helium cryostat. The phase transition between the high-temperature (4 × 1) and the low-temperature (8 × 2) phase was observed at a temperature of 125 K.

## Data analysis

One- and two-dimensional fast Fourier transforms of super Gaussian windowed time traces yield the frequency content of our data. The employed window function is:

$$F_{filt,t} = exp(-\ (\frac{(t-t_{shift})^2}{2\sigma_t^2})^3)$$

The employed parameters are:

| | Fig. 3b | Fig. 3c and Fig. 4d | Fig. 3d and Fig. 4e (two-dimensional) |
|---|---|---|---|
| $t_{shift}$ | 8 ps | 4 ps | 4 ps |
| $\sigma_t$ | 12 ps | 4.5 ps | 4.5 ps |

A minimum value is subtracted from the total switching yield in each delay trace to reduce the influence of low-frequency components. In Fig. 3c and Fig. 4d, the Fourier amplitude is normalized to the DC component for each individual switch pulse delay trace. In Fig. 3d and Fig. 4e, the normalization is performed with respect to the DC component of the two-dimensional Fourier transform.

**Density Functional Theory**

We performed DFT simulations within the local-density approximation (LDA)[48] as implemented within the Vienna Ab Initio Simulation Package (VASP)[49]. The electronic structure is described by projector-augmented wave potentials with a plane-wave basis set limited to a cut-off energy of 250 eV. A 2 × 8 × 1 Monkhorst-Pack mesh was used to sample the Brillouin zone. The surface was modeled using periodic boundary conditions and a slab with three bilayers of silicon. Si dangling bonds at the bottom layer were saturated with hydrogen.

In order to compute the two-dimensional potential energy surface (2D PES), which connects the (8×2) and (4×1) states along the amplitude modes, we first compute the direct transformation vector between both structures. In a second step, this vector is partitioned into shear and rotation contributions. For this purpose, we performed a constrained geometry relaxation of the (4×1) atomic structure in the (8×2) unit cell. The interatomic distances between the outermost In atoms along the wire direction were constrained to equal the respective distances found in the trimers of the (8×2) hexagon structure. The transformation vector connecting this trimerized structure to the (8×2) hexagon structure is equal to the shear deformation. Analogously, the transformation vector connecting the trimerized state to the ideal (4×1) state is the rotation deformation.

Constrained DFT [50] was used to obtain excited-state 2D PESs and perform *ab initio* molecular dynamics (AIMD) within the adiabatic approximation. In order to describe specific excited electronic configurations, the respective electronic states were identified numerically by means of their orbital character and their occupation numbers were held constant for the 2D PES calculations and AIMD simulations. The electronic occupation of the relevant band structure segments is based on the findings of previous studies. ARPES experiments revealed that the photoexcited electrons are strongly delocalized throughout the surface Brillouin zone, whereas photoholes are localized at the zone boundary [21]. The same authors demonstrated that the inclusion of electronic self-energy effects beyond standard DFT is necessary to recover the preferential confinement of photoholes in agreement with the ARPES experiments. However, self-energy corrected AIMD simulations are computationally intractable. Ref. [21] therefore compensated the misalignment in the valence state energies on an *ad hoc* basis, constraining occupation numbers in the AIMD simulations in accordance with Ref. [17], so that photoholes are localized to the zone-boundary and the zone-center valence states are fully occupied. For the purpose of the present study, we follow the suggested numerical treatment, incorporating hot-electron populations of photoexcited charge carriers in the lowest conduction states at the zone center and highest valence states at the zone boundary.

In order to clearly illustrate the vibrational coherence of the nuclear motion and the ballistic nature of the transition, we performed the AIMD simulations without a thermostat in the microcanonical ensemble at T = 0 K (only the coarse-grained molecular dynamics include a thermostat, cf. next section). The structural response to the excited electronic occupations increased the temperature only minimally, even upon successful switching into the (4x1) phase. We observed a maximum temperature increase of 6 K. Specific individual multi-pulse sequences are modeled in AIMD by adjusting the electronic occupations over the course of the AIMD at the arrival times of the respective pulses.

**Coarse-Grained Molecular Dynamics**

Modeling the full dynamics in all vibrational modes within a time-varying multidimensional PES derived from first principles is computationally intractable. We have therefore confined the parameter space to the shear and rotation deformations of the lattice and effectively incorporate coupling to other degrees of freedom using fluctuation-dissipation theory and experimentally determined damping constants [51,52]. The two-dimensional PES is calculated using constrained DFT in discrete steps of electronic occupation and interpolated to the momentary charge carrier density. The time-dependent electronic excitation is approximated by a Gaussian temporal profile with experimentally determined pulse width (see Methods: Ultrafast LEED and optical set-up), followed by an exponential decay. The intensity of the electronic excitation is chosen to match the experimentally observed relative switching yield (preparation pulses: 0.18 $e^-$/band/u.c., switch pulse: 0.36 $e^-$/band/u.c.). Electronic (0.25/ps) and phononic (shear: 0.19/ps rot: 0.17/ps) damping constants are estimated from time trace fits. An averaged switching yield into the (4×1) local minimum is determined by stochastic sampling of 2000 trajectories. Particle velocity rescaling, based on the measured base temperature ($T_c$=60 K) incorporates effective coupling to remaining other modes [52]. As a consequence of the adiabatic approximation, the resulting phonon frequencies are overestimated by about 20% and thus normalized to AIMD simulations.

## Extended Data Figures

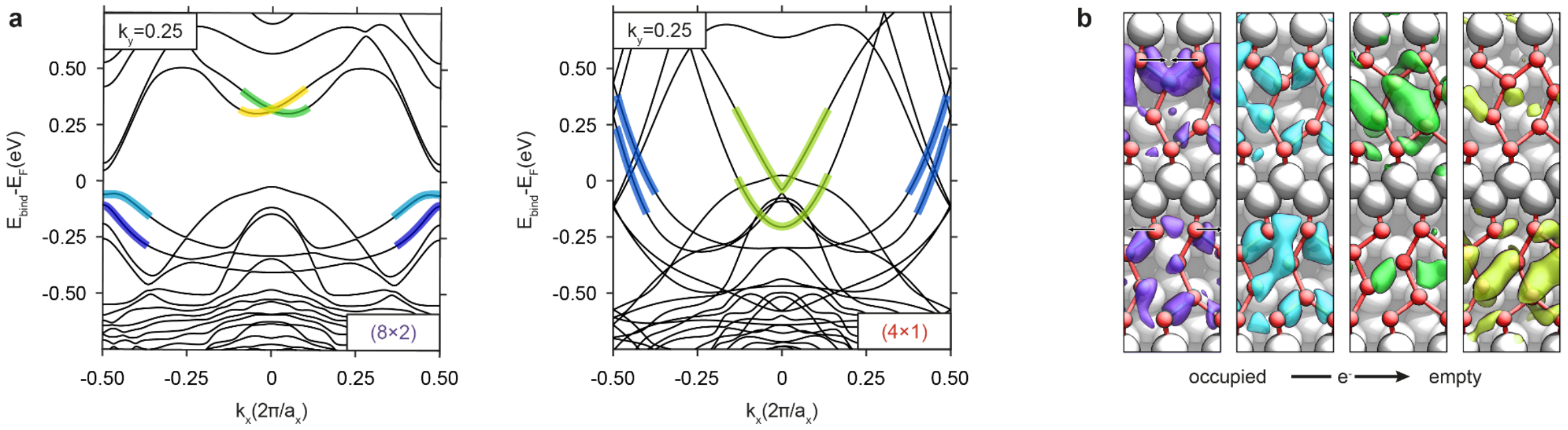

**Fig. S1: a**, Calculated electronic band structure of the Si(111)(8×2)-In (left) and Si(111)(4×1)-In (right) phase. $k_x$ and $a_x$ denote the reciprocal and real space lattice vectors in wire direction. **b**, Isodensity plots of wave function square moduli for the coloured states in a, with an isovalue of 0.005 $e^-/Å^3$.